# Observing Atomic Collapse Resonances in Artificial Nuclei on Graphene

Yang Wang[1,2,†], Dillon Wong[1,2,†], Andrey V. Shytov[3], Victor W. Brar[1,2], Sangkook Choi[1], Qiong Wu[1,2], Hsin-Zon Tsai[1], William Regan[1,2], Alex Zettl[1,2], Roland K. Kawakami[5], Steven G. Louie[1,2], Leonid S. Levitov[4], Michael F. Crommie[1,2*]

[1] Department of Physics, University of California at Berkeley, Berkeley CA, 94720, United States.
[2] Materials Science Division, Lawrence Berkeley National Laboratory, Berkeley CA, 94720, United States.
[3] School of Physics, University of Exeter, Stocker Road, Exeter EX4 4QL, United Kingdom.
[4] Department of Physics, Massachusetts Institute of Technology, Cambridge MA, 02139, United States.
[5] Department of Physics and Astronomy, University of California at Riverside, Riverside CA, 92521, United States

*Correspondence should be addressed to M.F.C.
[†] These authors contributed equally to this work

**One Sentence Summary:** Atomic collapse states are observed near artificial supercritical nuclei fabricated on graphene.

**Abstract:** Relativistic quantum mechanics predicts that when the charge of a super-heavy atomic nucleus surpasses a certain threshold, the resulting strong Coulomb field causes an unusual "atomic collapse" state which exhibits an electron wave function component that falls toward the nucleus as well as a positron component that escapes to infinity. In graphene, where charge carriers behave as massless relativistic particles, it has been predicted that highly charged impurities should exhibit resonances corresponding to these atomic collapse states. We have observed the formation of such resonances around artificial nuclei (clusters of charged calcium dimers) fabricated on gated graphene devices via atomic manipulation with a scanning tunneling microscope (STM). The energy and spatial dependence of the atomic collapse state measured using STM revealed unexpected behavior when it is occupied by electrons.

It has long been predicted that relativistic quantum effects can render atoms unstable when the charge of a nucleus ($Z$) exceeds a certain critical value $Z_c$ (*1, 2*). In this "supercritical" regime, the previously stable atomic bound states are predicted to "dive" into the positron continuum and acquire a finite lifetime (*3*). Quite unlike the standard Bohr-state orbitals seen in atomic physics and semiconductor impurity systems, such supercritical quantum eigenstates represent the quantization of a semi-classical trajectory having an electron-like component spiraling inward toward the nucleus (hence the term "atomic collapse") before spiraling back out and coupling to a positron-like component that propagates away from the nucleus (*1, 3-5*).

Some efforts to observe atomic collapse states have been made through colliding heavy-ions and searching for the positron created when a "collapsing" electron is torn from the vacuum, but these results remain ambiguous due to the enormously large $Z_c \sim 170$ that is required (*6, 7*). Hopes to realize atomic collapse states rose when theory predicted that highly charged impurities in graphene could display the same atomic collapse physics as atomic nuclei, but with holes in the graphene playing the role of positrons (*8-10*). Since charge carriers in graphene behave as relativistic particles with a large fine structure constant value (*11-13*), the supercritical charge threshold is substantially lower than for atoms, with $Z_c$ on the order of unity (*8-10*). The atomic collapse state near a Coulomb impurity in graphene should appear abruptly as the impurity charge is increased above $Z_c$ and manifest as a spatially extended electronic resonance whose energy lies just below the Dirac point. Such resonances correspond to the electron-like part of the atomic collapse wavefunction (in contrast to the far field measurements of out-going positrons performed in heavy-ion collision experiments) (*8-*

*10*). Recently subcritical Coulomb behavior (i.e., where $Z < Z_c$ and atomic collapse resonances are absent) was observed for charged impurities in graphene (*14*), but the observation of atomic collapse around supercritical impurities has remained elusive due to the difficulty of producing highly charged impurities.

Here we report the observation of supercritical Coulomb behavior in atomically-fabricated "artificial nuclei" assembled on the surface of a gated graphene device. These tunable charge centers were synthesized by pushing together ionized calcium (Ca) impurities using the tip of an STM, thus allowing creation of supercritical Coulomb potentials from subcritical charge elements. STM spectroscopy was used to observe the emergence of atomic-collapse electronic states extending further than 10nm from the center of artificial nuclei in the supercritical regime ($Z > Z_c$). By tuning the graphene Fermi level ($E_F$) via electrostatic gating we observed behavior in this state that was unanticipated by theory (*8-10*).

We transferred chemical-vapor-deposition (CVD) grown graphene (*15*) onto a boron nitride (BN) flake placed on a $SiO_2$/Si surface (the doped Si substrate comprised a gate electrode). The BN substrate reduced the charge inhomogeneity of graphene (*16-18*) sufficiently that we could observe the intrinsic electronic behavior of pristine graphene adjacent to charged impurities and artificial nuclei. We used Ca dimers as charge building blocks for constructing supercritical artificial nuclei on graphene because they were more easily manipulated with an STM tip than Ca monomers. The dimers formed spontaneously by allowing the monomers to diffuse on the surface by warming to $16 \pm 3$ K for 1 to 2 mins [details of the assembly and characterization are available as supplemental online materials (Fig. S1) (*19*)].

The effect of a charged Ca dimer on nearby Dirac fermions in graphene was determined by measuring both the electron-like and hole-like local density of states (LDOS) near a single dimer through STM $dI/dV$ spectroscopy performed at different lateral distances from the center of a single Ca dimer (Fig.1A). These spectra exhibited a ~130-meV wide gap-like feature at $E_F$ caused by phonon-assisted inelastic tunneling, and an additional minimum around $V_s$ = +0.23V associated with the Dirac point (*20*). These spectra also display a systematic electron-hole asymmetry in that the LDOS intensity of states above the Dirac point increased as the tip neared the dimer while the LDOS of states below the Dirac point correspondingly decreased. This asymmetric behavior is characteristic of subcritical impurities and indicates that Ca dimers were positively charged on graphene at this doping level (*14*). Such electron-hole asymmetry was observed for Ca dimers at different gate voltages in the range -60V $\leq V_g \leq$ +30V [see supplement Fig. S6 (*19*)] and no signature of charge state instability was seen, indicating that Ca dimers were charge-stable within our measurement conditions.

Artificial nuclei containing up to five Ca dimers were assembled by using atomic manipulation techniques to push individual dimers into small clusters (see insets to Fig. 1, A to E). $dI/dV$ spectra measured with the STM tip held over the bare graphene near these artificial nuclei revealed the emergence of a resonance in the electronic LDOS (i.e., a quasi-bound state) as the number of dimers in a cluster was increased from one to five (Fig. 1, A to E). Spectra acquired near 2-dimer clusters (Fig. 1B) displayed greater electron-hole asymmetry than single dimer spectra, as well as an extra oscillation in LDOS at high energies above the Dirac point. Spectra taken near 3-dimer clusters (Fig. 1C) exhibited even stronger electron-hole asymmetry, and the oscillation in LDOS began here to coalesce into a resonance-like structure near Vs= +0.30V. For four dimer clusters, the intensity of the resonance feature increased markedly and moved down in energy to the Dirac point (Fig 1D). For the 5-dimer cluster, the resonance shifted below the Dirac point (Fig. 1E). The formation of this resonance (or quasi-bound state) as nuclear charge was increased is the "smoking gun" for atomic collapse, as we will shortly describe.

In order to better characterize the spatial dependence of the quasi-bound state that appears for high-$Z$ artificial nuclei, we performed *dI/dV* mapping at the resonance energy around a 5-dimer cluster (Fig. 2). The resonance intensity displayed a highly symmetric distribution around the 5-dimer cluster despite a marked asymmetry in the arrangement of dimers in the cluster. The intensity of the resonance extended outward from the cluster center to a distance greater than 10 nm. Thus, the resonance was derived from charge carriers moving in the pristine graphene that were spatially separate from the dimers making up the artificial nucleus.

The behavior of the quasi-bound (atomic collapse) state observed for high-$Z$ artificial nuclei depended on whether it was occupied by electrons or empty, as can be seen in the gate-dependent spectra of Fig. 3 which were all measured 3.7 nm from the center of a 5-dimer cluster. Red arrows in Fig. 3 indicate the energy of the atomic collapse state at each back-gate voltage, while black arrows indicate the energy of the Dirac point extracted by measuring *dI/dV* spectra on graphene far away (~20 nm) from the cluster at each back-gate voltage. In the p-doped regime (-60 V $\leq$ $V_g$ $\leq$ -15 V) the quasi-bound state shifted lower in energy (relative to $E_F$) as more electrons were added to the system, similar to the shift of the Dirac point except that the resonance transited from above to below the Dirac point as the p-doping was decreased. In the n-doped regime (0V

≤ Vg ≤ +30V), the resonance shifted below $E_F$ and its behavior changed dramatically. Here the resonance intensity decreased and essentially disappeared, although a small rise remained in the *dI/dV* just above the Dirac point.

The quasi-bound state that develops as the Ca-cluster charge increases is the atomic-collapse eigenstate of a supercritical charged nucleus. This explanation is supported by Dirac equation-based calculations of the expected LDOS for positively charged nuclei with different charge numbers (*8-10, 14*), which we compare to our experimental spectra. These *dI/dV* simulations assume a 2D continuum Dirac model for undoped graphene in the presence of a Coulomb potential. A boundary condition at short distance from the nucleus must be chosen to account for the finite size of the clusters. We thus set $V(r)$ to the constant value $V(r_0)$ for $r < r_0$, a similar condition as used in previous theoretical simulations of supercritical nuclei in vacuum (*1, 2*). We fixed the value of $r_0$ to 1nm, which matches the geometry of our clusters. We found that varying $r_0$ over the range 0.5 nm ≤ $r_0$ ≤ 1.5 nm did not markedly affect the quality of the fits to our data [simulations using other $r_0$ values are included in Fig. S4 (*19*)]. The only essential fitting parameter in our simulation was the ratio of the cluster charge to the critical value, $Z/Z_c$. Here we define $Z$ as the "effective charge", i.e. the screened cluster charge as seen by Dirac electrons. The value $Z$ absorbs the effects of intrinsic screening due to graphene band polarization (*14*) and screening due to the substrate. The critical value in this case is $Z_c = \hbar v_F/(2e^2) \sim 0.25$, where $v_F$ is the Fermi velocity [this $Z_c$ expression differs slightly from the one used for isolated 3D atoms, see supplement for a detailed discussion (*19*)]. Additional perturbations from lifetime broadening and electron-phonon coupling are accounted for as in (*14*) and (*21*).

Our data suggests that clusters with just one or two Ca dimers are in the subcritical regime, and so $Z/Z_c$ here is determined by matching the amplitude of the electron-hole asymmetry between simulation and experiment according to the method described in ref. (*14*). Our clusters comprised of 3 or more dimers are either transitioning into (3 dimers) or have fully entered (4 and 5 dimers) the supercritical regime, as evidenced by the sharpness and energy location of the quasi-bound state resonance (*8, 9*). For these clusters, $Z/Z_c$ is determined by matching the quasi-bound state resonance energy between the simulation and experiment. The resulting simulated $dI/dV$ spectra from the Dirac equation are plotted in Figs. 1, F to J, next to the corresponding experimental data. The main features seen in our experimental data are well reproduced by the Dirac equation simulations. In particular, our Dirac equation simulations reproduce the increase in electron-hole asymmetry observed in the subcritical regime as the number of dimers increases from one to two, as well as the emergence of the supercritical atomic collapse state as the number of dimers is increased from three to five. The $Z/Z_c$ values extracted from our Dirac equation fits are $Z/Z_c = 0.5 \pm 0.1$ (one dimer), $0.9 \pm 0.1$ (two dimers), $1.4 \pm 0.2$ (three dimers), $1.8 \pm 0.2$ (four dimers), and $2.2 \pm 0.2$ (five dimers) [further details of the simulations, including bare LDOS, can be found in the supplement].

In order to check that the magnitude of $Z/Z_c$ extracted for Ca dimers from our Dirac equation fits is physically reasonable, we additionally performed a completely separate calculation of the charge state expected for a Ca dimer adsorbed to graphene using *ab initio* density functional theory (DFT). This separate calculation (which had no fitting parameters) yielded a single-dimer charge ratio of $Z/Z_c = 0.6 \pm 0.3$ [only the single-dimer case was calculated using DFT, see supplement *(19)*]. This is in agreement with the value $Z/Z_c = 0.5 \pm 0.1$ obtained via our Dirac equation simulations, and thus lends further support to our overall interpretation of the data.

We considered a possible alternative explanation for the quasi-bound state resonance seen for our Ca dimer clusters involving multiple scattering of electrons (*22*) between different Ca dimers within a cluster. This mechanism, however, can be excluded because the distance between Ca dimers is $d \sim 2$ nm, which would result in a peak energy measured from the Dirac point of $E = \hbar v_F k \sim \hbar v_F(\pi/d) \sim 1$ eV. This value exceeds by more than an order of magnitude the energy of the state observed for Ca clusters on graphene (at most 0.05eV). The large spatial extent (Fig. 2) of the quasi-bound state resonance shows that it arises from an extended Coulomb field and thus also rules out the possibility that it is simply caused by local impurity hybridization with underlying carbon atoms. The radially symmetric distribution observed is consistent with the predicted behavior of the atomic collapse state having lowest angular momentum ($j=1/2$) and lowest energy (*8, 9*). In principle, an infinite number of resonances (corresponding to different principle quantum numbers) should appear in the supercritical regime. However, because their energies follow an exponential sequence and their spatial extent is inversely proportional to their energy measured from the Dirac point (*8, 9*), only the lowest energy state should be detectable within our experimental resolution (screening suppresses all states with sufficiently large spatial extent).

The strong doping dependence of the intensity of the atomic collapse state (Fig. 3) can be partially explained as a screening effect. In the p-doped regime the atomic collapse state is experimentally observed to shift to lower energy with respect to the Dirac point for decreased p-doping. This can be explained by free-electron-like screening (*23, 24*) since reduced screening should result in an increase of the effective charge of the artificial nucleus. The n-doped regime, however, shows a completely different behavior. Here the amplitude of the atomic collapse state essentially disappears as it moves beneath $E_F$ (i.e., as it becomes occupied with charge carriers). We believe that this behavior arises from internal correlation among the quasi-localized charge carriers that inhabit the atomic collapse state (*25*) (only one out of the four electron states allowed due to spin and valley degeneracy can be occupied at any given time). Interaction between these electrons reduces the single-particle spectral function as measured in STM spectroscopy, but the precise mechanism for this reduction remains to be determined.

**References and Notes:**
1. W. Greiner, B. Muller, J. Rafelski, *Quantum Electrodynamics of Strong Fields* (Springer-Verlag, Berlin, 1985), pp.
2. I. Pomeranchuk, Y. Smorodinsky. About the Energy Levels of Systems with Z > 137. *J. Phys. USSR* **9**, 97 (1945).
3. Y. B. Zeldovic, V. S. Popov. Electronic Structure of Superheavy Atoms. *Sov. Phys. Usp.* **14**, 673 (1972).
4. C. G. Darwin. On some orbits of an electron. *Philosophical Magazine Series 6* **25**, 201 (1913).
5. T. H. Boyer. Unfamiliar trajectories for a relativistic particle in a Kepler or Coulomb potential. *American Journal of Physics* **72**, 992 (2004).
6. T. Cowan *et al.* Anomalous Positron Peaks from Supercritical Collision Systems. *Physical Review Letters* **54**, 1761 (1985).
7. J. Schweppe *et al.* Observation of a Peak Structure in Positron Spectra from U+Cm Collisions. *Physical Review Letters* **51**, 2261 (1983).
8. V. M. Pereira, J. Nilsson, A. H. Castro Neto. Coulomb Impurity Problem in Graphene. *Physical Review Letters* **99**, 166802 (2007).
9. A. V. Shytov, M. I. Katsnelson, L. S. Levitov. Atomic Collapse and Quasi-Rydberg States in Graphene. *Physical Review Letters* **99**, 246802 (2007).
10. A. V. Shytov, M. I. Katsnelson, L. S. Levitov. Vacuum Polarization and Screening of Supercritical Impurities in Graphene. *Physical Review Letters* **99**, 236801 (2007).
11. A. H. Castro Neto, F. Guinea, N. M. R. Peres, K. S. Novoselov, A. K. Geim. The electronic properties of graphene. *Reviews of Modern Physics* **81**, 109 (2009).
12. K. S. Novoselov *et al.* Two-Dimensional Gas of Massless Dirac Fermions in Graphene. *Nature* **438**, 197 (2005).
13. Y. Zhang, Y. W. Tan, H. L. Stormer, P. Kim. Experimental observation of the quantum Hall effect and Berry's phase in graphene. *Nature* **438**, 201 (2005).
14. Y. Wang *et al.* Mapping Dirac quasiparticles near a single Coulomb impurity on graphene. *Nat Phys* **8**, 653 (2012).
15. X. Li *et al.* Large-Area Synthesis of High-Quality and Uniform Graphene Films on Copper Foils. *Science* **324**, 1312 (June 5, 2009, 2009).
16. C. R. Dean *et al.* Boron nitride substrates for high-quality graphene electronics. *Nat Nano* **5**, 722 (2010).
17. R. Decker *et al.* Local Electronic Properties of Graphene on a BN Substrate via Scanning Tunneling Microscopy. *Nano Letters* **11**, 2291 (2011).
18. J. Xue *et al.* Scanning tunnelling microscopy and spectroscopy of ultra-flat graphene on hexagonal boron nitride. *Nat Mater* **10**, 282 (2011).
19. Supplementary materials are available on *Science* online.
20. Y. Zhang *et al.* Giant phonon-induced conductance in scanning tunnelling spectroscopy of gate-tunable graphene. *Nature Physics* **4**, 627 (2008).
21. V. W. Brar *et al.* Observation of Carrier-Density-Dependent Many-Body Effects in Graphene via Tunneling Spectroscopy. *Physical Review Letters* **104**, 036805 (2010).
22. E. J. Heller, M. F. Crommie, C. P. Lutz, D. M. Eigler. Scattering and absorption of surface electron waves in quantum corrals. *Nature* **369**, 464 (1994).

23. T. Ando. Screening Effect and Impurity Scattering in Monolayer Graphene. *The Physical Society of Japan* **75**, 074716 (2006).
24. E. H. Hwang, S. Das Sarma. Dielectric function, screening, and plasmons in two-dimensional graphene. *Physical Review B* **75**, 205418 (2007).
25. V. N. Kotov, B. Uchoa, V. M. Pereira, F. Guinea, A. H. Castro Neto. Electron-Electron Interactions in Graphene: Current Status and Perspectives. *Reviews of Modern Physics* **84**, 1067 (2012).
26. J. Li, W.-D. Schneider, R. Berndt. Local density of states from spectroscopic scanning-tunneling-microscope images: Ag(111). *Physical Review B* **56**, 7656 (1997).
27. C. Wittneven, R. Dombrowski, M. Morgenstern, R. Wiesendanger. Scattering States of Ionized Dopants Probed by Low Temperature Scanning Tunneling Spectroscopy. *Physical Review Letters* **81**, 5616 (1998).

**Acknowledgements**: Research supported by the Office of Naval Research Multidisciplinary University Research Initiative (MURI) award no. N00014-09-1-1066 (graphene device preparation, characterization and imaging), by the Director, Office of Science, Office of Basic Energy Sciences of the US Department of Energy under contract no. DE-AC02-05CH11231 (STM instrumentation development and spectroscopy), by the National Science Foundation award nos. EEC-0832819 (dI/dV simulations) and DMR10-1006184 (DFT calculations), and by EPSRC grant EP/G036101/1 (Dirac eqn. calculations). Computational resources were provided by DOE at the LBNL NERSC facility.

**Figure Captions:**

**Fig. 1.** Evolution of charged impurity clusters from subcritical to supercritical regime. **(A-E)** *dI/dV* spectra measured at different distances from the center of Ca-dimer clusters (i.e., artificial nuclei) composed of 1-5 dimers (initial tunneling parameters: $V_s$ = -0.50V, I = 60pA, AC modulation $V_{rms}$ = 6mV). "Center" here is defined as the average coordinate of dimers within a cluster. All spectra were acquired at the same back-gate voltage ($V_g$ = -30V) and each was normalized by a different constant factor to account for exponential changes in conductivity due to location-dependent tip-height changes (*14, 26, 27*). Insets: STM topographs of atomically fabricated Ca-dimer clusters. **(F-J)** Theoretical normalized dI/dV spectra (obtained from Dirac equation) for graphene at same distances from dimer clusters as in **(A-E)** for the nuclear charges $Z/Z_c$ of **(F)** 0.5 (1-dimer cluster), **(G)** 0.9 (2-dimer cluster), **(H)** 1.4 (3-dimer cluster), **(I)** 1.8 (4-dimer cluster), and **(J)** 2.2 (5-dimer cluster). Black dashed lines indicate Dirac point, red arrows indicate atomic collapse state observed both in experiment and in Dirac simulation.

**Fig. 2.** *dI/dV* map near a 5 Ca-dimer cluster at the energy of the supercritical quasi-bound state resonance, as marked by red arrow in Fig. 1E (tunneling parameters: $V_s$ = +0.20V, I = 15pA, $V_g$ = -30V). The Ca dimers appear as slightly darker disks near the center of the *dI/dV* map.

**Fig. 3.** Dependence of atomic collapse state on doping. Gate dependent spectra acquired at a lateral distance of 3.7 nm from the center of a five-dimer Ca cluster (initial tunneling parameters: $V_s$ = -0.50V, I = 60pA, AC modulation $V_{rms}$ = 6 mV). Curves are shifted

vertically for clarity. Red arrows indicate energy of the atomic collapse state at each back-gate voltage. Black arrows indicate energy of the Dirac point extracted by measuring the *dI/dV* spectrum on graphene far from the cluster center (~20 nm) at each back-gate voltage. The atomic collapse state intensity is quenched in the n-doped regime.

**Supplementary Materials:**
Materials and Methods
Supplementary Text
Figures S1-S10
References (*28-37*)

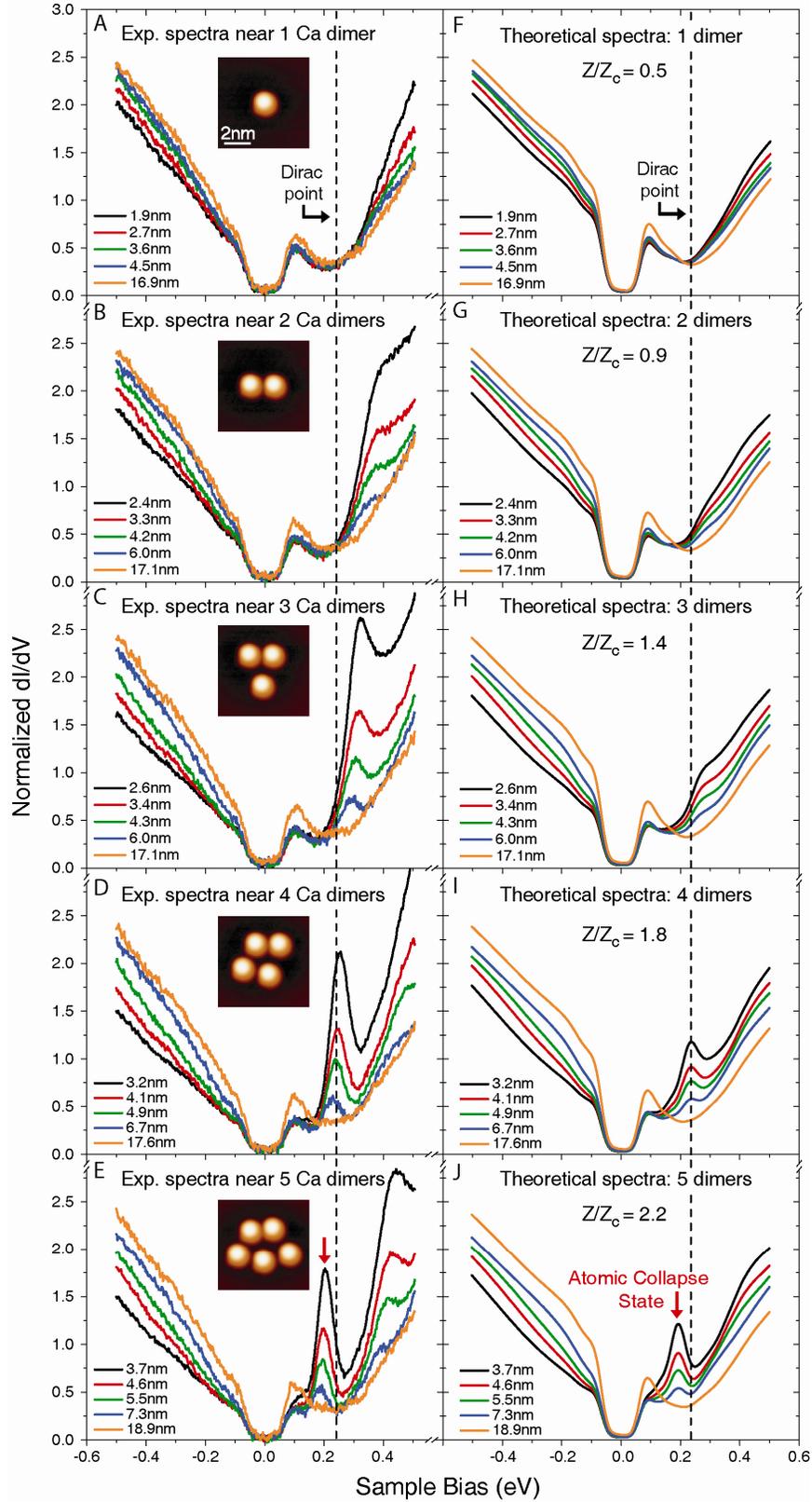

**Fig. 1**

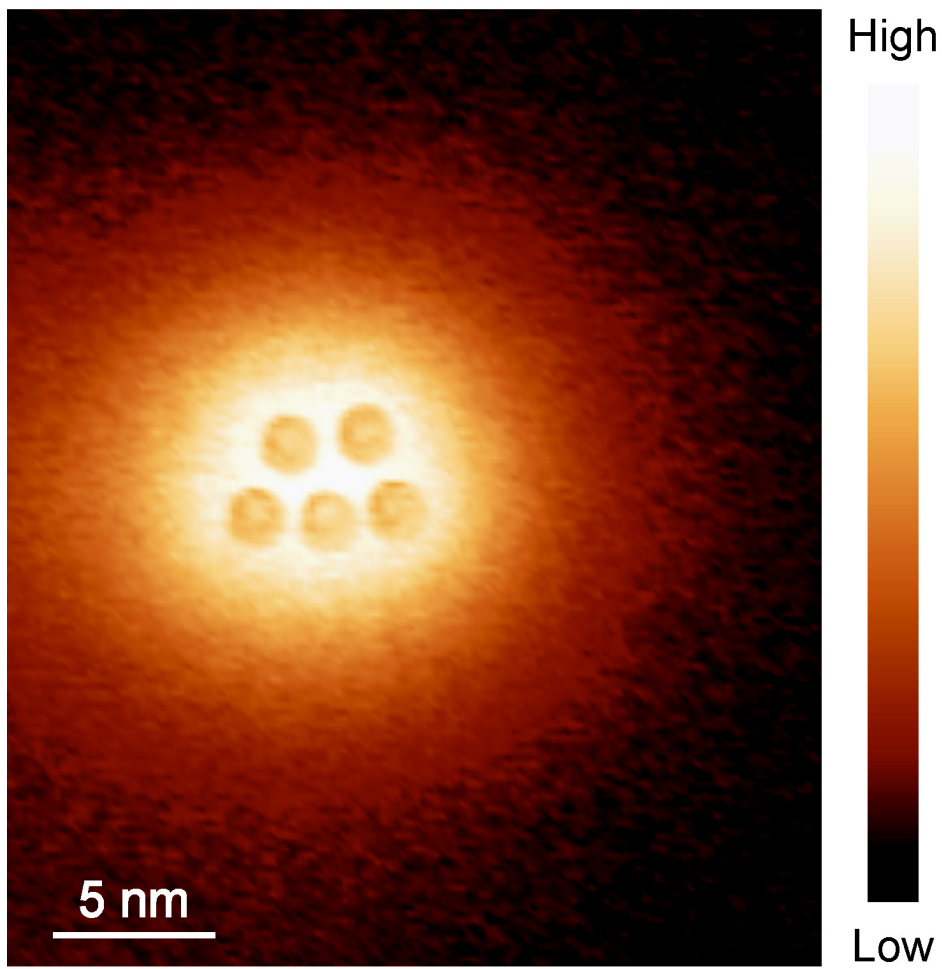

**Fig. 2**

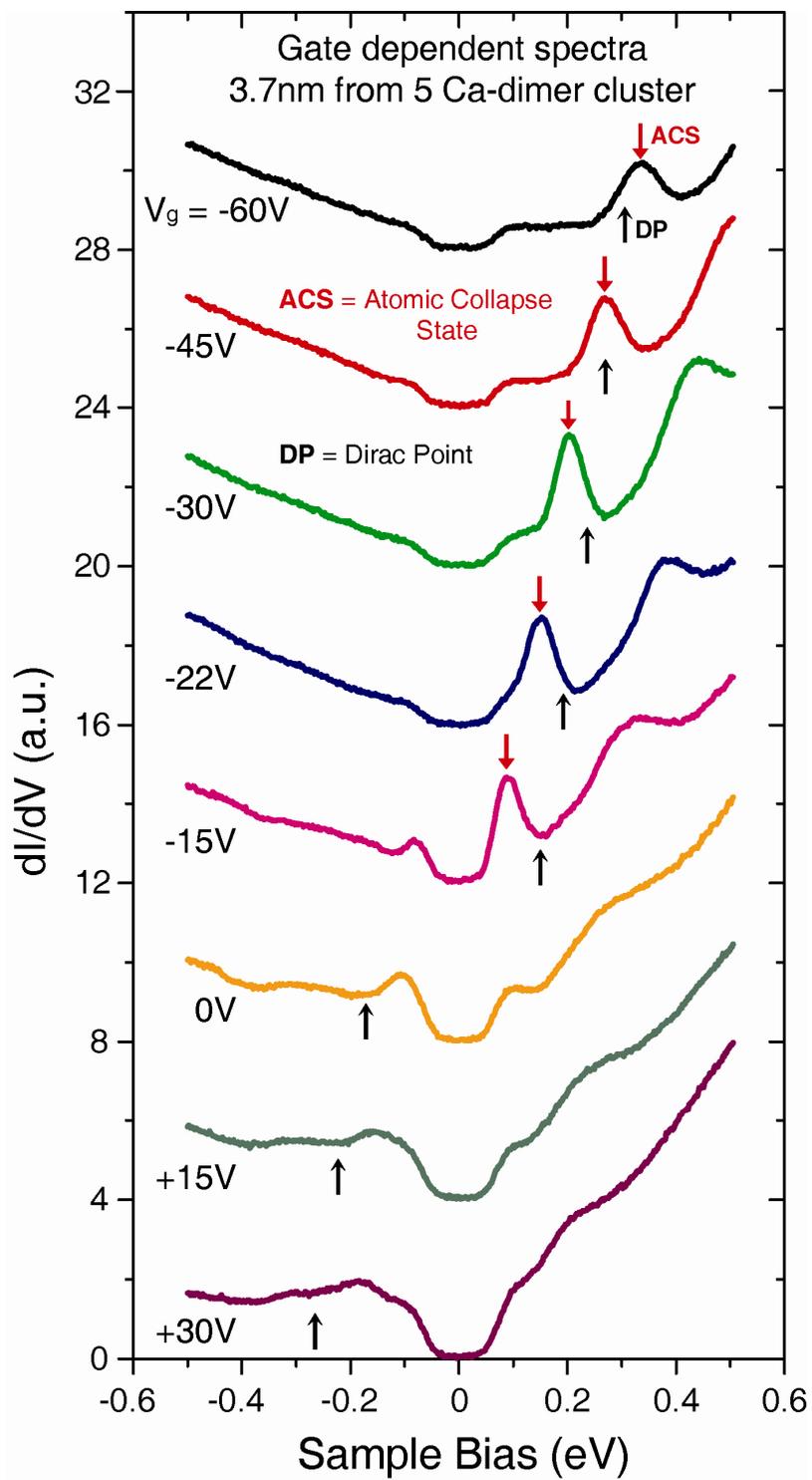

**Fig. 3**